\documentclass[11pt]{article}
\usepackage{amsmath,amssymb}
\usepackage{amsfonts}
\usepackage{eucal}
\title{\leavevmode\vadjust{\vskip -5mm}
\textbf{The phase symmetry  \\
of general relativity}}
\author{\sc Serhii Samokhvalov\thanks{{\bf e}-{\it mail}: serh.samokhval@gmail.com}\\
 \tabaddress{Dniprovsk State Technical University, Ukraine}}

\date{ ***\\
} 

\def\tabaddress#1{{\small\it\begin{tabular}[t]{c}#1
\\[1.2ex]\end{tabular}}}
\def\<#1>{\langle#1\rangle}
\def\beq{\begin{equation}}
\def\eeq{\end{equation}}
\def\bea{\begin{eqnarray}}
\def\eea{\end{eqnarray}}
\def\beann{\begin{eqnarray*}}
\def\eeann{\end{eqnarray*}}
\def\ben{\begin{enumerate}}
\def\een{\end{enumerate}}
\def\qed{\ifvmode\removelastskip\fi
{\unskip\nobreak\hfil\penalty50\hbox{}\nobreak\hfil \hbox{\vrule
height1.2ex width1.2ex}\parfillskip=0pt \finalhyphendemerits=0
\par\smallskip}}
\def\texthook{\vrule height 0pt depth 0.4pt width 3.5pt
          \vrule height 5pt depth 0.4pt \kern 3pt}
\def\scripthook{\vrule height 0pt depth 0.2pt width 1.5pt
                \vrule height 3pt depth 0.2pt width 0.2pt \kern 1pt}

\begin{document}

\maketitle

\thispagestyle{empty}

\begin{abstract}

It is shown that the general relativity has a one-parameter compact symmetry and this symmetry is analogous to the phase symmetry of quantum mechanics
\end{abstract}

%%%%%%%%%%%%%%%%%%%%%%%%%%%%%%%%%%%%%%%%%%%%%%%%%%%%%%%%%%%%%%%%
\clearpage
%%%%%%%%%%%%%%%%%%%%%%%%%%%%%%%%%%%%%%%%%%%%%%%%%%%%%%%%%%%%%%%%
%\tableofcontents
%%%%%%%%%%%%%%%%%%%%%%%%%%%%%%%%%%%%%%%%%%%%%%%%%%%%%%%%%%%%%%%%
%\clearpage
%%%%%%%%%%%%%%%%%%%%%%%%%%%%%%%%%%%%%%%%%%%%%%%%%%%%%%%%%%%%%%%%

%%%%%%%%%%%%%%%%%%%%%%%%%%%%%%%%%%%%%%%%%%%%%%%%%%%%%%%%%%%%%%%%
%\section{Introduction}
%%%%%%%%%%%%%%%%%%%%%%%%%%%%%%%%%%%%%%%%%%%%%%%%%%%%%%%%%%%%%%%%

\section{Introduction}

The two main general properties of matter that were discovered in the creation of quantum mechanics, namely, the discreteness (quantization) of certain physical quantities, as well as the wave nature of motion, are associated with the existence of fundamental compact symmetry - phase symmetry of quantum mechanics. From the point of view of the theory of symmetry, we can say that the discovery of quantum-mechanical laws was the discovery of the fundamental phase symmetry inherent in the laws of motion of matter.

However, a certain compact symmetry is inherent in other wave phenomena, in particular the electromagnetic field in the vacuum - the so-called Rainich-Heaviside symmetry \cite{1}. The study of this symmetry (which by analogy with quantum mechanics we also called phase symmetry) was performed in \cite{2}, where several interesting properties of this symmetry were revealed. In particular, using the fact that this symmetry is a consequence of the discrete symmetry of Maxwell's equations in vacuum at the transformation of the Hodge duality for the electromagnetic field tensor, an explicitly phase-symmetric so-called "vector Lagrangian" of electrodynamics similar to Dirac's Lagrangian was constructed. This Lagrangian has the interesting property that the quantity conserved due to phase symmetry (when applying the Neother theorem to this Lagrangian) is proportional to the energy-momentum tensor of the electromagnetic field. 

In this paper, it is shown that the gravitational field in vacuum is also characterized by a certain compact one-parameter symmetry, which is a generalization to the case of the gravitational field of Rainich-Heaviside symmetry of the electromagnetic field and which we will also call the phase symmetry. To reveal this symmetry, general relativity (GR) is considered in an orthogonal frame, which corresponds to the natural interpretation of the gravitational field as the gauge field of the translation group \cite{3}. It is shown that the energy-momentum tensor of the gravitational field is phase-symmetric. An explicitly phase-symmetric "vector Lagrangian" of gravity is constructed and it is shown that similarly to the case of an electromagnetic field the Noether current of such Lagrangian, which associated with phase symmetry, is proportional to the energy-momentum tensor of the gravitational field.

Greek indices belong to the coordinate basis, and Latin - to the frame basis. The transition between them (indices replacement) is carried out using mutually inverted matrices $h_m^\mu$ and $h^m_\mu$. A point is placed in the place of the lowered or raised index (where this may lead to misunderstanding).

\section{Phase symmetry of classical electrodynamics}

We first present the main provisions concerning the phase symmetry of classical electrodynamics \cite{2}, presenting them in a form convenient for our purposes generalization to the case of gravity.

The electromagnetic field strength tensor $F_{\mu \nu}$ is expressed in terms of potentials $A_\mu$ as follows
\[
F_{\mu \nu}=\partial_\mu A_\nu-\partial_\nu A_\mu
\]
where $ \partial_\mu:= \partial  / \partial x^\mu$. Jacobi identity written for $F_{\mu \nu}$
\begin{equation}\label{eq1}
\partial_\rho F_{\mu \nu}+cicle(\rho \mu \nu)=0
\end{equation}
can be rewritten as:
\begin{equation}\label{eq2}
\partial_\nu \widetilde{F}^{\mu \nu}=0,
\end{equation}
where
\begin{equation}\label{eq3}
\widetilde{F}^{\mu \nu}:=\frac{1}{2}\varepsilon^{\mu\nu\rho\tau}F_{\rho\tau}
\end{equation}
is the tensor, dual according to Hodge to the tensor $F_{\mu \nu}$ and $\varepsilon^{\mu \nu \rho \tau}$ is absolutely antisymmetric tensor of Minkowski space. Equation (\ref{eq2}) (or (\ref{eq1})), with three-dimensional reduction, decomposes into the first pair of Maxwell equations.

The second pair of Maxwell equations in four-dimensional form in the absence of sources
\begin{equation}\label{eq4}
\partial_\nu F^{\mu \nu}=0
\end{equation}
are found from Lagrangian
\begin{equation}\label{eq5}
L^M=-\frac{1}{16\pi} F_{\mu \nu} F^{\mu \nu}
\end{equation}
(here the speed of light $c=1$).

Infinitesimal transformations with parameter $\delta \varphi$
\begin{equation}\label{eq6}
\delta F^{\mu \nu}= \widetilde{F}^{\mu \nu} \delta \varphi
\end{equation}
and the transformations derived from it (taking into account formula (\ref{eq3}))
\begin{equation}\label{eq7}
\delta \widetilde{F}^{\mu \nu}= -F^{\mu \nu} \delta \varphi
\end{equation}
keep the system of Maxwell equations in vacuum (\ref{eq2}), (\ref{eq4}) invariant, generalize the Rainich-Heaviside transformations and are called phase transformations of classical electrodynamics \cite{2}.

At the phase transformations, the Lagrangian $L^M$ (\ref{eq5}) obtains the term:
\[
\delta L^M=-\frac{1}{8\pi} F_{\mu \nu}\delta F^{\mu \nu}=-\frac{1}{8\pi} F_{\mu \nu} \widetilde{F}^{\mu \nu} \delta \varphi=-\frac{1}{4\pi}\partial_\mu A_\nu \widetilde{F}^{\mu \nu} \delta \varphi=
\]
\[
-\frac{1}{4\pi}[\partial_\mu (A_\nu \widetilde{F}^{\mu \nu})- (A_\nu \partial_\mu \widetilde{F}^{\mu \nu}] \delta \varphi,
\]
which due to identity (\ref{eq2}) (the first pair of Maxwell equations) is reduced to divergence. Therefore, the phase transformations (\ref{eq6}) are the symmetry of electrodynamics generalized in the sense of \cite{4} and preserves (due to (\ref{eq2})) second pair of Maxwell equations (\ref{eq4}).

In \cite{2} it was shown that there is a so-called "vector Lagrangian" of electrodynamics
\begin{equation}\label{eq8}
L_\rho^D=\frac{l}{8\pi}(\widetilde{F}_{\rho\nu}\partial_\mu F^{\nu\mu}-F_{\rho\nu}\partial_\mu \widetilde{F}^{\nu\mu}),
\end{equation}
similar to Dirac's Lagrangian ($ l $ is length dimension constant), which has the following properties:

\noindent - it is symmetric with respect to the transformation of the Hodge duality (\ref{eq3}) and explicitly phase-symmetric (invariant with respect to transformations (\ref{eq6}), (\ref{eq7}));

\noindent - if we take the tensors $F^{\mu \nu}$ and $ \widetilde{F}^{\mu \nu} $ as independent variables in it, then when they are varied, both pairs of Maxwell equations (\ref{eq2}) and (\ref{eq4}) are obtained as Lagrange equations;

\noindent - phase transformations (\ref{eq6}), (\ref{eq7}) lead for Lagrangian $L_\rho ^D$ (8) to Noether current
\begin{equation}\label{eq9}
J_\rho^\mu=-\frac{l}{8\pi}(F^{\sigma\mu}F_{\sigma\rho}+\widetilde{F}^{\sigma\mu}\widetilde{F}_{\sigma\rho})=
-\frac{l}{4\pi}(F^{\sigma\mu}F_{\sigma\rho}-\frac{1}{4}\delta_\rho^\mu F^{\sigma \tau}F_{\sigma \tau})
\end{equation}
which coincides (with accuracy to multiplier $l$) with the energy-momentum tensor of the electromagnetic field. We note also the invariance of the tensor $J_\rho^\mu$ with respect to the duality transformation (\ref{eq3}) and phase transformations (\ref{eq6}), (\ref{eq7}).

\section{General relativity in an orthogonal frame}

Symmetry, similar to that considered above, is inherent in the theory of gravity, if it is presented in the frame form \cite{5}, \cite{3}. In this case, the potentials of the gravitational field are the components of (pseudo)orthonormal frame fields $h_\mu^m$, and the strength of the gravitational field is described by anholonomic coefficients
\[
F_{\mu \nu}^m=\partial_\nu h_\mu^m-\partial_\mu h_\nu^m.
\]

In this section we discuss the basic relations of the theory of gravity in an orthogonal frame, which are used below.

Due to the rotor structure of the tensor $F_{\mu \nu}^m$, the Jacobi identity also holds for it
\begin{equation}\label{eq10}
\partial_\rho F_{\mu \nu}^m+cicle(\rho \mu \nu)=0,
\end{equation}
which can be given the form:
\begin{equation}\label{eq11}
\nabla_\nu \widetilde{F}^{m \mu\nu}=\frac{1}{h}\partial_\nu (h \widetilde{F}^{m \mu\nu})=0,
\end{equation}
where $\nabla_\nu$ is a covariant derivative in Riemannian space with metric $g_{\mu \nu}=h_\mu ^m h_\nu ^n \eta_{mn}$, $\eta_{mn}$ is a metric of Minkowski space, $h=\det\,h_\mu ^m$ and
\begin{equation}\label{eq12}
\widetilde{F}^{m \mu\nu}:=\frac{1}{2}\varepsilon^{\mu\nu\rho\tau}F_{\rho\tau}^m
\end{equation}
is the Hodge dual tensor to the gravitational field strength tensor $F_{\mu \nu}^m$. Here $\varepsilon^{\mu \nu \rho \tau}$ is the absolutely antisymmetric tensor of four-dimensional Riemannian space in curvilinear coordinates. Identity (\ref{eq11}) (or (\ref{eq10})) in a Riemannian space with a metric $g_{\mu \nu}$ is equivalent to the so-called cyclic identity:
\[
R_{\;\:\rho\mu\nu}^m+cicle(\rho\mu\nu)=0
\]
and in the gravity theory plays a role similar to the first pair of equations of Maxwell's electrodynamics.

Einstein equations in vacuum
\[
G_m ^\mu :=R_m ^\mu-\frac{1}{2}h_m ^\mu R=0
\]
derived from truncated Hilbert's Lagrangian (Möller's Lagrangian):
\begin{equation}\label{eq13}
L^H=-\frac{1}{2}R-\nabla_\sigma R^\sigma=\frac{1}{2}\delta_{mn}^{s\:p}\:\omega_{\;\:\: s\,l}^m \:\omega_{\; pn}^l
\end{equation}
(we assume that Einstein's gravitational constant $\kappa=1$). Here $ G_m^\mu$ is the Einstein tensor, $ R_m^\mu$ is the Ricci tensor, and $R$ is the scalar curvature, $\delta_{mn}^{s\,p}:=\delta_{m}^{s}\delta_{n}^{p}-\delta_{m}^{p}\delta_{n}^{s}$ - alternator,
\[
\omega_{nlk}^{\,\cdot}:=\frac{1}{2}(F_{\;lnk}^{\,\cdot}+F_{\;nlk}^{\,\cdot}-F_{\;knl}^{\,\cdot})
\]
- Ricci rotation coefficients and $R_p:=\nabla_\sigma h_p^\sigma=F_{sp}^s=\omega_{\;sp}^s$.

Using the gravitational field induction tensor (superpotential)
\begin{equation}\label{eq14}
B_m^{\mu\nu}:=\omega_{\;m\,\cdot}^{\mu\ \:\nu}-\delta_{mp}^{\mu\nu}R^p
\end{equation}
allows us write down Lagrangian $L^H$ (\ref{eq13}) in a form similar to the form of Maxwell's Lagrangian $L^M$ (\ref{eq5}):
\begin{equation}\label{eq15}
L^H=\frac{1}{4}F_{\mu\nu}^m B_m^{\mu\nu},
\end{equation}
and Einstein equation in the form:
\begin{equation}\label{eq16}
G_m ^\mu =-\nabla_\nu B_m^{\mu\nu}-t_m ^\mu=0,
\end{equation}
where
\begin{equation}\label{eq17}
t_m ^\mu=-\frac{1}{h}\partial_{h_m ^\mu}(h L^H)=B_n^{\nu\mu}F_{\nu m}^n-h_m ^\mu L^H
\end{equation}
is the energy-momentum tensor of the gravitational field in the theory of gravity in an orthogonal frame \cite{3}, \cite{4}.

\section{Duality and phase symmetry of general relativity}

Einstein equations (\ref{eq16}) are nonlinear, are much more complex than the equations of electrodynamics (\ref{eq4}) (second pair of Maxwell equations) and are not invariant with respect to transformations
\[
\delta F_{\;\;\;\;\cdot\:\cdot}^{m \mu\nu}=\widetilde{F}^{m \mu\nu}\delta \varphi
\]
similar to transformations (\ref{eq6}) which constructed on the basis of the Hodge duality transformation (\ref{eq12}) for the gravitational field strength tensor. But, as will be shown in this section, Einstein equations (\ref{eq16}) are invariant with respect to compact one-parameter transformations, which are based on more refined discrete duality transformation for the gravitational field strength tensor than the Hodge duality transformation (\ref{eq12}).

In order to detect such transformations, we reverse formula (\ref{eq14}), i.e., we express the tensor of the gravitational field $F_{\mu \nu}^m$ through the tensor of its induction $B_m^{\mu \nu }$. First of all, note that $B_s^{s \nu }=-2R^\nu$, whence it follows
\[
\omega_{\;m\,\cdot}^{\mu\ \:\nu}=B_m^{\mu\nu}-\frac{1}{2}\delta_{mp}^{\mu\nu}B_s^{sp}
\]
Thus, since $ F_{\mu \nu}^m =\delta_{\mu\nu}^{k\,l}\omega_{\;\:k\,l}^m$, the result is:
\begin{equation}\label{eq18}
F_{\mu\nu}^m=\delta_{\mu\nu}^{k\,l}(B_{\:k\,\;l}^{m\;\cdot}-\frac{1}{2}\delta_{k\:p}^{mn}\eta_{n\,l}B_s^{sp}).
\end{equation}
Using this formula, Lagrangian $L^H$ (15) can be written in the form of a quadratic form with respect to the gravitational field induction tensor:
\begin{equation}\label{eq19}
L^H=\frac{1}{2}B_{\:n\,s}^{m\;\cdot}B_{\:m}^{n\;s}-\frac{1}{4}B_{\:m\,s}^{m\;\,\cdot}B_{\:n}^{n\;s}.
\end{equation}

Let us now consider the discrete transformation of the gravitational field induction tensor
\begin{equation}\label{eq20}
B_m^{\mu\nu}\rightarrow \widehat{B}_m^{\mu\nu}:=\widetilde{F}_m^{\,\cdot \,\mu\nu},
\end{equation}
which, according to (\ref{eq18}), generates the following transformation of the strength tensor:
\begin{equation}\label{eq21}
F_{\mu\nu}^m\rightarrow\widehat{F}_{\mu\nu}^m=\delta_{\mu\nu}^{k\,l}(\widetilde{F}_{\;\;\;\;k\,l}^{m\;\cdot\;\cdot}-
\frac{1}{2}\delta_{k\:p}^{mn}\eta_{n\,l}\widetilde{F}_{\:s}^{\,\cdot\,sp}).
\end{equation}
A unique property of the transformation (\ref{eq21}), which is tested directly, is the fact that for a dual Hodge tensor of the gravitational field strength, it leads to the transformation:
\begin{equation}\label{eq22}
\widetilde{F}_m^{\,\cdot \,\mu\nu}\rightarrow \widetilde {\widehat{F}}{}_m^{\,\cdot \,\mu\nu}=-B_m^{\mu\nu},
\end{equation}
which indicates that this transformation, as well as the transformation of the Hodge duality in a four-dimensional pseudo-Riemannian space, has the properties of the imaginary unit, i.e. its repeated performance only leads to multiplication by -1. This is ensured the fact that a continuous infinitesimal transformations with a parameter $\delta \varphi$
\begin{equation}\label{eq23}
\delta F_{\mu\nu}^m=\widehat{F}_{\mu\nu}^m \delta\varphi,
\end{equation}
built on the basis of duality transformation (\ref{eq21}), is compact, therefore, like the phase transformations of quantum mechanics, associated with certain wave properties of the now gravitational field in the vacuum. The infinitesimal transformations (\ref{eq23}) in terms of the dual tensors $B_m^{\mu \nu}$ and $\widetilde{F}_m^{\,\cdot \,\mu\nu}$ are written as follows:
\begin{equation}\label{eq24}
\delta B_m^{\mu \nu}=\widetilde{F}_m^{\,\cdot \,\mu\nu} \delta\varphi,
\end{equation}
\begin{equation}\label{eq25}
\delta \widetilde{F}_m^{\,\cdot \,\mu\nu}=-B_m^{\mu \nu} \delta\varphi
\end{equation}
and in the finite version takes the form:
\[
{B'}_m^{\mu \nu}=B_m^{\mu \nu} \cos \varphi+\widetilde{F}_m^{\,\cdot \,\mu\nu} \sin \varphi,
\]
\[
{\widetilde{F}{'}}_m^{\,\cdot \,\mu\nu}=-B_m^{\mu \nu} \sin \varphi+\widetilde{F}_m^{\,\cdot \,\mu\nu} \cos \varphi.
\]

Property (\ref{eq22}) occurs precisely because the gravitational field induction tensor is given by expression (\ref{eq14}), which follows from the truncated Hilbert's Lagrangian (\ref{eq13}) of general relativity. This gives grounds to call the quantities $F_ {\mu \nu}^m$ and $\widehat{F}_ {\mu \nu}^m $ (as well as $B_m^{\mu \nu}$ and $ \widetilde{F}_m^{\mu \nu}$) as \textbf{GR-dual quantities}, and transformation (\ref{eq21}) (as well as transformations (\ref{eq20}) and (\ref{eq22}) agreed with them) as \textbf{transformations of GR-duality}.

The following shows that transformations (\ref{eq24}), (\ref{eq25}), by virtue of identity (\ref{eq11}), preserve Einstein equations in the frame variables (\ref{eq16}), and therefore are the symmetry of Einstein's theory of gravity in the (pseudo)orthonormal frame, which we will call \textbf{the phase symmetry of general relativity}. Indeed, due to the quadraticity of the Lagrangian $L^H$ with respect to $B_m^{\mu \nu}$ (see formula (\ref{eq19})) and using formula (\ref{eq24}), we find that during the phase transformations Lagrangian $ L^H $ (\ref{eq15}) obtains the term:
\[
\delta L^H=\frac{1}{2} F_{\mu \nu}^M \delta B_m^{\mu \nu}=\frac{1}{2} F_{\mu \nu}^m \widetilde{F}_m^{\,\cdot \,\mu \nu} \delta \varphi=-\partial_\mu h_\nu^m \widetilde{F}_m^{\,\cdot \,\mu \nu} \delta \varphi=
\]
\[
-\frac{1}{h}[\partial_\mu (h \widetilde{F}_m^{\,\cdot \, \mu m})-h_\nu^m \partial_\mu (h \widetilde{F}_m^{\,\cdot \, \mu \nu})] \delta \varphi=[\nabla_\mu \widetilde{F}_m^{\,\cdot \, \mu m}-h_\nu^m \nabla_\mu \widetilde{F}_m^{\,\cdot \, \mu \nu}]\delta \varphi.
\]
When performing identity (\ref{eq11}) this term reduced to covariant divergence that ensures the invariance of equations (\ref{eq16}), which follow from the Lagrangian $L^H$ (Einstein equations in the orthogonal frame), at the phase transformations (\ref{eq24}) (or (\ref{eq23})).

	The method of the above proof of the phase symmetry of Einstein equations shows that the arbitrary quadratic (relative to the strength tensor $F_{\mu \nu}^m$) frame theory of gravity is invariant with respect to transformations (\ref{eq24}) with its induction tensor $B_m^{\mu \nu}$, which is determined by the Lagrangian of the theory $L=\frac{1}{4}F_{\mu\nu}^m B_m^{\mu\nu}$. The compactness of transformations (\ref{eq24}), which provided by property (\ref{eq22}) and therefore corresponds to a certain wave process, demonstrated by Einstein's theory of gravity with Lagrangian (\ref{eq13}), that distinguishes it from others. But Einstein's frame theory of gravity is unambiguously fixed among quadratic theories also by the requirement of local Lorentz invariance $\Lambda^g$. The fact that the $\Lambda^g$-invariant theory is also phase-symmetric probably has a deep basis, which may be related to the property of the absence of volume deviation in Einstein's space-time during geodetic transfer \cite{6}.

\section{Consequences of phase symmetry of general relativity and "vector Lagrangian"}

The above-proven invariance of Einstein equations with respect to phase transformations (\ref{eq24}) under condition (\ref{eq11}) when applied directly to equation (\ref{eq16}) gives
\[
\delta G_m ^\mu =-\nabla_\nu \delta B_m^{\mu\nu}-\delta t_m ^\mu=-\nabla_\nu \widetilde{F}_m^{\,\cdot \,\mu\nu} \delta\varphi-\delta t_m ^\mu=-\delta t_m ^\mu=0,
\]
therefore, the energy-momentum tensor of the gravitational field $t_m^\mu$ (\ref{eq17}) is phase-invariant. This is easy to verify directly if the tensor $t_m^\mu$ is represented as:
\begin{equation}\label{eq26}
t_m ^\mu=\frac{1}{2}(B_s^{n \mu}F_{n m}^s+\widetilde{F}_s^{\cdot \,n \mu} \widetilde{B}_{\,\cdot \,nm}^s),
\end{equation}
or in the variables $\widetilde{F}_m^{\cdot \,\mu \nu}$ and $B_m^{\mu \nu}$ in the form:
\begin{equation}\label{eq27}
t_m ^\mu=\frac{1}{4}\varepsilon_{mnkl}(B_s^{n \mu}\widetilde{F}^{skl}-\widetilde{F}_s^{\cdot \,n \mu} B_{\,\cdot}^{skl}),
\end{equation}
and apply transformations (\ref{eq24}), (\ref{eq25}). Really,
\[
\delta t_m ^\mu=\frac{1}{4}\varepsilon_{mnkl}(\delta B_s^{n \mu}\widetilde{F}^{skl}+B_s^{n \mu}\delta \widetilde{F}^{skl}-
\delta \widetilde{F}_s^{\cdot \,n \mu} B_{\,\cdot}^{skl}-\widetilde{F}_s^{\cdot \,n \mu} \delta B_{\,\cdot}^{skl})=
\]
\[
\frac{1}{4}\varepsilon_{mnkl}(\widetilde{F}_s^{\cdot \,n \mu}\widetilde{F}^{skl}-B_s^{n \mu} B_{\,\cdot}^{skl}+
 B_s^{n \mu} B_{\,\cdot}^{skl}-\widetilde{F}_s^{\cdot \,n \mu}  \widetilde{F}^{skl}) \delta \varphi=0.
\]

Equations (\ref{eq11}) in the frame theory of gravity play the role of the first pair of Maxwell equations in electrodynamics, but in the general case are not preserved at phase transformations. Really,
\begin{equation}\label{eq28}
\nabla_\nu \delta \widetilde{F}_m^{\cdot \, \mu\nu}=-\nabla_\nu \delta B_m^{\mu\nu} \delta \varphi=t_m^\mu \delta \varphi.
\end{equation}
The last equation in (\ref{eq28}) is obtained under the condition that Einstein equations (\ref{eq16}) hold. Therefore, the phase symmetry of the complete system of equations of the gravitational field (\ref{eq11}) and (\ref{eq16}) occurs only under the condition
\begin{equation}\label{eq29}
t_m^\mu =0.
\end{equation}
In this case, the equations of the gravitational field become linear with respect to the variables $\widetilde{F}_m^{\,\cdot \,\mu\nu}$ and $B_m^{\mu \nu}$:
\begin{equation}\label{eq30}
\nabla_\nu \widetilde{F}_m^{\,\cdot \,\mu\nu}=0,\qquad \nabla_\nu B_m^{\mu \nu}=0,
\end{equation}
moreover, the transformation of the GR-duality (\ref{eq20}), (\ref{eq22}) translate these equations into one another, and thus become the symmetry of the complete system of gravitational equations (30). Note that the energy-momentum tensor of the gravitational field is also symmetric with respect to the transformation of duality (\ref{eq20}), (\ref{eq22}):
\[
\widehat{t}_m^{\,\mu}=t_m^{\,\mu},
\]
that immediately follows from expression (\ref{eq27}). Thus, condition (\ref{eq29}) is dual and phase symmetric.

The energy-momentum tensor of the gravitational field $t_m^\mu$ is only a coordinate tensor (vector) and in the case of $\Lambda^g$-transformations associated with changes of reference frames \cite{7} is transformed according to the non-tensor law \cite{8}. Thus, the question arises under what conditions, given the $\Lambda^g$-invariance of Einstein's theory of gravity, we can find such $\Lambda^g$-gauging (such reference frame), for which $t_m^\mu=0$ and, thus, there is a phase symmetry. For example, in \cite{9} it is shown that condition (\ref{eq29}) is satisfied in a free-falling reference frame in the Schwarzschild field and it is assumed that in an arbitrary gravitational field in a free-falling reference frame the energy-momentum tensor of the gravitational field must be zero due to the equivalence principle. If so, then the requirement of phase symmetry of gravity theory is equivalent to the requirement of choosing free-falling reference frames. This question, of course, requires further study.

	In the frame theory of gravity, as well as in electrodynamics, it is possible to construct a "vector Lagrangian" like Dirac's Lagrangian:
\begin{equation}\label{eq31}
L_\rho ^V=\frac{l}{2}(\widetilde{B}_{\,\rho\;\nu}^{m\,\cdot}\nabla_\mu B_m^{\nu \mu}-F_{\rho\nu}^m \nabla_\mu \widetilde{F}_m^{\,\cdot \,\nu\mu}),
\end{equation}
which, like Lagrangian (\ref{eq8}), has the properties:

\noindent - it is symmetric with respect to the GR-duality transformation (\ref{eq21}) and explicitly phase symmetric (invariant with respect to transformations (\ref{eq24}), (\ref{eq25}));

\noindent - if we take the tensor densities $h B_m^{\mu \nu}$ and $h \widetilde{F}_m^{\,\cdot \,\mu\nu}$ as independent variables, then when they are varied, from Lagrangian $L_\rho^V$ (\ref{eq31}) both tensor equations of the gravitational field (\ref{eq30}) are obtained as Lagrange equations - cyclic identity and Einstein equation at the condition (\ref{eq29});

\noindent - for the Lagrangian $L_\rho^V$, phase transformations (\ref{eq24}), (\ref{eq25}) lead to the Noether current
\begin{equation}\label{eq32}
J_\rho^\mu=\frac{l}{2}(B_s^{n \mu}F_{n \rho}^s+\widetilde{F}_s^{\cdot \,n \mu} \widetilde{B}_{\,\cdot \,n \rho}^s)=l \, t_m^\mu h_\rho^m,
\end{equation}
which (up to the multiplier $l$) coincides with the energy-momentum tensor of the gravitational field (\ref{eq26}), written in the coordinate frame.
%\vskip 3mm

%\textbf{Conclusions}
\section{Discussion and conclusions}

It is established that Einstein's gravity theory in the orthogonal frame, like quantum mechanics and electrodynamics, demonstrates (under certain conditions) the presence of compact phase symmetry, and we found the duality transformation that provides this phenomenon.

The conditions of phase symmetry of the gravitational field equations are considered and the possible connection of the phase symmetry of gravity with the principle of equivalence is indicated.

A phase-symmetric "vector Lagrangian" of the gravitational field is constructed and it is shown that the consequence of its phase symmetry is the conservation law of energy-momentum of the gravitational field.

However, a number of questions regarding phase symmetry in gravity remain unresolved. This is, first of all, the geometric nature of phase symmetry and its connection with the local Loretz invariance of the theory as well as with the principle of equivalence.

In addition, the meaning of the so-called "vector Lagrangian" in the theory of gravity remains unclear, as well as its mysterious property, that the consequence of the phase symmetry of the "vector Lagrangian" is the conservation of energy-momentum of the gravitational field, while for truncated Hilbert's Lagrangian energy-momentum is conserved due to the generalized translational invariance of the theory \cite{4}.

All of these questions require further study, which may shed light on the connection between gravity and quantum mechanics.

%%%%%%%%%%%%%%%%%%%%%%%%%%%%%%%%%%%%%%%%%%%%%%%%%%%%%%%%%%%%%%%%

\end{document}